\documentclass[11pt,a4paper,twoside]{article} 
\usepackage{amsmath}
\usepackage{amsfonts}
\usepackage{color}
\usepackage{rotating}
\usepackage{graphicx}
\usepackage[T1]{fontenc} 
\usepackage{bm}
\usepackage{hyperref}
\usepackage{version}

\usepackage[english]{babel}
\usepackage{pdflscape}
\usepackage{float}

\numberwithin{equation}{section}
\begin{document}
\vspace{1cm}

\vspace{1cm}

\noindent
{\bf
{\large 3+1D Massless Weyl spinors from bosonic scalar-tensor duality
}}

\vspace{.5cm}
\hrule

\vspace{1cm}

\noindent

{\large{Andrea Amoretti$^{a,}$\footnote{\tt andrea.amoretti@ge.infn.it},
Alessandro Braggio$^{b,}$\footnote{\tt alessandro.braggio@spin.cnr.it }, 
Giacomo Caruso$^{a,}$\footnote{\tt giacomo.caruso@ge.infn.it }, 
Nicola Maggiore$^{a,}$\footnote{\tt nicola.maggiore@ge.infn.it },
Nicodemo Magnoli$^{a,}$\footnote{\tt nicodemo.magnoli@ge.infn.it }
\\[1cm]}}

{\small{{}$^a$  Dipartimento di Fisica, Universit\`a di Genova,
via Dodecaneso 33, I-16146, \\Genova, Italy\\and\\INFN, Sezione di Genova\\
\medskip
{}$^b$ CNR-SPIN, Via Dodecaneso 33, 16146, Genova, Italy}}

\vspace{1cm}
\noindent

{\tt Abstract~:}
We consider the fermionization of a bosonic free theory characterized by the $3+1D$ 
scalar - tensor duality. This duality can be interpreted as the dimensional reduction, via a 
planar boundary, of the 
$4+1D$ topological BF theory. In this model, adopting the Sommerfield tomographic 
representation of quantized bosonic fields, we explicitly build a fermionic operator and 
its associated Klein factor such that it satisfies the correct anticommutation relations.
Interestingly, we demonstrate that this operator satisfies the massless Dirac equation and that it can be identified with a $3+1D$ Weyl spinor. 
Finally, as an explicit example, we write the integrated charge density in terms of the tomographic transformed bosonic degrees of freedom.
\vfill\noindent
{\footnotesize 
{\tt PACS Nos:} 
03.70.+k	Theory of quantized fields;
11.10.-z	Field theory;
03.65.Pm Dirac equation;
05.30.-d Fermi-Dirac statistics.

}
\newpage
\begin{small}
\end{small}

\setcounter{footnote}{0}

\section{Introduction}

Dualities play an important role in various branches of theoretical physics since long 
time\cite{Savit:1980sa}. Perhaps the first example is the electromagnetic duality between 
electric and magnetic fields in Maxwell equations. More recently, but still two decades ago, 
the Seiberg--Witten duality \cite{Seiberg:1994rs} relating the weak and strong coupling regime of 
N=2 Super Yang--Mills theories induced a set of dualities between various string 
theories \cite{Giveon:1994fu}. 
In particular, fruitful applications have been found and exploited in condensed matter physics, where the 
AdS/CFT correspondence \cite{Witten:1998qj,Polchinski:2010hw}, as a realization of the gauge/gravity 
duality, has found recent important developments \cite{Sachdev:2010ch}. 

The common feature of dualities, is that of relating different physics, which otherwise would not 
be related one with each other. The example we are dealing with in this paper is the duality 
which permits to build fermionic degrees of freedom (d.o.f.) out of bosonic ones.

There has been a huge activity in 
the opposite direction, that is the formulation in terms of bosons of a (possibly interacting) 
fermionic theory. The problem has been firstly solved in 1+1 
dimensions
 \cite{Luttinger63,MattisLieb65,Mattis75,Coleman75,Mandelstam75,Haldane81,Shankar:1995rm,ShoellerDelft98}, 
but is 
still an open question in higher dimensions and many efforts are continuously devoted toward this 
goal \cite{Luther:1978cy,Burgess:1994np, Chan:2012nb}. 

Less popular is the reverse process of fermionization where fermionic 
fields  (operators) are obtained from bosonic counterparts. Fermionization has 
been considered in spin liquid states and heavy fermions \cite{Tsvelik92,Coleman93}. In higher 
dimensions, using the tomographic representation of quantized fields presented 
in \cite{Sommerfield:1982qf}, the problem was solved
both in $2+1D$ \cite{Aratyn:1984jz} and $3+1D$ \cite{Aratyn:1983bg}. 
There, the starting point to proceed to the fermionization is the assumption of a duality 
relation, introduced somehow by hand, between some bosonic fields.

A stronger motivation for considering the fermionization recently rised, again, from condensed matter, 
and it can be summarized as follows. New states of matter have been predicted (and discovered), 
called topological insulators (TI) \cite{HasanKane10,QiZhang11}, for which the low energy 
physics seem to 
be captured by a class of Schwarz-type 
topological quantum field theory: the BF models \cite{Birmingham:1991ty,Cho:2010rk,Blasi12}. 
Being topological, those theories acquires local d.o.f. only on the boundary. 

For what concerns the 
$2+1D$ TI, the abelian BF model may be easily rephrased in terms of  
two Chern-Simons theories with opposite coupling constants  \cite{Blasi12,Guadagnini90}. On 
its $1+1D$ planar boundary the action depends on two scalar fields satisfying a duality relation 
and it may be rewritten in terms of two counter-propagating bosonic chiral modes. The duality makes also 
possible to fermionize the bosonic d.o.f. into two counter-propagating 
chiral electronic modes connected by time 
reversal symmetry: the helical Luttinger liquid.\footnote{The T symmetry require that when one of the 
two modes have only one spin component, the other necessarily has opposite one.} 
 
For what concerns the $3+1D$ TI, the BF model has a $2+1D$ bosonic boundary action that depends only on 
a scalar and a vector field, satisfying the duality relation which allows 
to pursue the fermionization procedure described in \cite{Aratyn:1984jz} inferring the 
presence of fermionic d.o.f. on the boundary, which is what is needed to describe the surface states 
of a topological insulator.  An alternative approach \cite{Qi08} for $3+1D$ TI is to consider the 
effective theories for the $4+1D$ TI,  where a Chern-Simon satisfying T symmetry may 
be considered, making then a dimensional reduction (compactification) to recover 
the $3+1D$ or the $2+1D$ TI bulk theories \cite{QiZhang11}.

In conclusion we know that duality relations in a bosonic theory could give rise to a fermionization 
procedure. A systematic theoretical framework to get bosonic boundary actions with consistent duality 
relations relating bosonic fields exists in any spacetime $d$-dimension. This is indeed 
represented by the BF topological models, which can be defined in generic $d$ spacetime dimensions, 
with a planar boundary. The crucial ``fermionizing'' duality relations are not imposed $ad\ hoc$, but turn 
out to be the most general boundary conditions for the bosonic fields. 

The result is that to any $p$-form in the bulk corresponds a $(p-1)$-form on the boundary. The $d=3+1D$ case
has been treated in \cite{Amoretti:2012hs}, where the duality relation involves a scalar and a vector 
field, and the fermionic construction has been done in \cite{Aratyn:1984jz}. 

Here we pursue the program in $d=4+1D$, where, starting from the BF model, we get the duality relation 
on the $3+1D$ planar boundary. In this case, the resulting bosonic fields are  a scalar and a tensor field, with no vector field.  
  
It is worth to mention that our construction gives Weyl spinors in $3+1D$, and it could be relevant in 
the recently discussed Weyl semimetal physics where, even if the bulk is gapless, still the boundary 
modes are topologically protected \cite{Burkov11,Xu11}.

The paper is organized as follows. In section 2 we describe and motivate the duality relation we start with, identifying the bosonic d.o.f. to be fermionized. In section 3 we implement the fermionization procedure. First we review some basic properties of the tomographic transform, then we apply this method to our system and finally, starting from the fermionic variable that we have obtained, we argue how to construct a Weyl spinor. In section 4, as an example, we relate the total charge with this expressed in terms of the tomographic transformed bosonic fields of our starting model. In section 5 we summarize our results.

\section{Duality and bosonic degrees of freedom} \label{dag}

\paragraph{Preliminaries} $\;$\\

We recall some of the results obtained in \cite{Amoretti:2013nv}, which represent our starting point. There, it has been shown the possibility to dimensionally reduce the abelian $4+1D$ BF model, described by the action
\begin{equation}
\label{actionbf5}
S=\frac{1}{2} \int d^5x \epsilon^{\alpha\beta\gamma\delta\eta} \partial_\alpha A_\beta B_{\gamma\delta\eta} 
\end{equation}
to a gapless bosonic theory on the $3+1D$ planar boundary $x_4=0$. In particular, it has been found that it is possible to parametrize the fields on the boundary in terms of a scalar potential $\Lambda$ and an antisymmetric tensor potential \mbox{$\phi_{\mu \nu}=-\phi_{\nu \mu}$}, with \mbox{$\mu, \; \nu...=\{0,1,2,3\}$}, as follows:
\begin{eqnarray}
&A_{\mu}=\partial_{\mu} \Lambda,\\
&B_{\nu \rho \sigma}=\partial_{[\mu} \phi_{\rho \sigma]},
\end{eqnarray}
where $[...]$ means antisymmetrization of indices. The dynamics on the boundary is completely determined by the $3+1D$ Lagrangian:
\begin{equation}
\label{lag5d4d}
\mathcal{L}=\frac{1}{2} \epsilon^{ijk} \partial_i \phi_{jk} \partial_t \Lambda+\frac{1}{2}\epsilon^{ijk}\partial_{i} \Lambda  \partial_t \phi_{jk}-(\epsilon^{ijk}\partial_i \phi_{jk})^2-\frac{1}{4} ( \partial_i \Lambda )^2,
\end{equation}
where $i, \;j...={1,2,3}$. The Lagrangian \eqref{lag5d4d} is invariant under the following gauge symmetries:
\begin{eqnarray}
\label{gaigi1}
\delta^{(1)}\Lambda= c \qquad \;\\ 
\label{gaigi2}
\delta^{(2)} \phi_{\mu \nu} = \partial_{[\mu} \alpha_{\nu]}
\end{eqnarray}
where $c$ is a constant and $\alpha_\mu$ is a $3+1D$ vector gauge parameter. 

The dimensional reduction on the plane $x_4=0$ induces an unique boundary condition which, written in terms of the boundary fields $\Lambda$ and $\phi_{\mu\nu}$, reads: 
\begin{equation}
\epsilon^{\mu\nu\rho\sigma}\partial_{\nu}\phi_{\rho\sigma}+\partial^{\mu}\Lambda=0. \label{duality}
\end{equation}
This boundary condition will play the role of the fermionizing duality relation which is the starting point of \cite{Aratyn:1984jz} and \cite{Aratyn:1983bg}. \\
Summarizing, we are dealing with a non-topological $3+1D$ field theory \eqref{lag5d4d} defined on a 
flat Minkoskian spacetime with metric $g_{\mu\nu}\equiv diag(-1,1,1,1)$. At this point, it is natural to ask 
which is the physics described by the 3+1D theory \eqref{lag5d4d} constrained by the duality relation \eqref{duality}. The main purpose of this paper is to answer this question.

\paragraph{The gauge choice and the independent degrees of freedom} $\;$\\

Since the boundary model \eqref{lag5d4d} is invariant under the gauge symmetries \eqref{gaigi1} and \eqref{gaigi2}, it is necessary to fix a gauge for the consistency of the theory.\\
To do this, we note that, differentiating \eqref{duality} with respect to $x_\mu$ we find that the scalar field $\Lambda$ is massless:
\begin{equation}
\Box \Lambda=0.
\end{equation}
Then, multiplying \eqref{duality} by $\epsilon_{\alpha\beta\gamma\mu}$ and differentiating with respect to $x_{\alpha}$, we obtain:
\begin{equation}
\Box\phi_{\beta\gamma} + \partial^{\alpha}\partial_{\beta}\phi_{\gamma\alpha} + \partial^{\alpha}\partial_{\gamma}\phi_{\alpha\beta}=0,  \label{asd}
\end{equation}
which is exactly the equation of motion which must be satisfied by a free massless tensor (see Appendix of \cite{Kalb:1974yc}). In particular an admissible gauge choice for $\phi_{\mu \nu}$ is
\begin{equation}
\phi_{0i}=
\partial^{j}\phi_{ij}=0. \label{gauge22}
\end{equation}
With this gauge choice it is easy to see that $\phi_{\mu \nu}$ can be parametrized in terms of a massless vector field $\xi^i$:
\begin{equation}
\phi^{jk}=\frac{1}{2}\epsilon^{ijk}\xi_i,
\end{equation}
with the condition that $ \xi_i$ is a longitudinal field:
\begin{equation}
\epsilon^{ijk}\partial_j\xi_k=0. \label{roto}
\end{equation}
Consequently, $\phi_{\mu \nu}$ has only one degree of freedom, according to the general rule of the Kalb-Ramond fields \cite{Kalb:1974yc}.

\section{The fermionization procedure}

We are dealing with a $3+1D$ Lagrangian \eqref{lag5d4d} which involves a scalar massless field $\Lambda$ and a longitudinal vector field $\xi^i$, related by the duality relation \eqref{duality}. 
In \cite{Cho:2010rk} the duality relation used in \cite{Aratyn:1984jz} between the boundary bosonic d.o.f. has been used to guess the existence of the fermionic d.o.f. of the $2+1D$ topological insulators. Inspired by this, in this section we apply the same method to the duality relation \eqref{duality}, to prove that, in the low energy limit, the $3+1D$ boundary model described by \eqref{lag5d4d} has fermionic excitations as well.\\
Preliminarly, we recall that the fermionization procedure relies on the tomographic transform presented in \cite{Sommerfield:1982qf}, which displays a known non-locality drawback \cite{Luther:1978cy}. Nevertheless this problem is not relevant in our case, since, as we said, we are dealing with a low energy effective bosonic field theory.

\subsection{The Tomographic Transform}

In this section we review some basic properties of the tomographic transform, which we conveniently use in the fermionization process. For more details on the tomographic representation we refer to \cite{Sommerfield:1982qf}.
\\
The basic ingredient of the tomographic transform in three spatial dimensions is the generalized function $ \delta'(y-\bm n \cdot \bm r)$, defined as:
\begin{equation}
 \delta'(y-\bm n \cdot \bm r) = \int_{-\infty}^{+\infty}{ dk\, k e^{iky} e^{-i \bm r \cdot k \bm n}}, \label{deltaprimo}
\end{equation}

where $\bm n$ is an angular variable, and $k$ is a scalar. From the definition \eqref{deltaprimo}, it is easy to prove that $ \delta'(y-\bm n \cdot \bm r)$ satisfies the completeness and the ortonormality relation:
\begin{gather}
\frac{1}{8\pi^2}\int{dy\,d^2 n\,\delta'(y-\bm n \cdot \bm r)\delta'(y-\bm n \cdot \bm r')}=\delta(\bm r -\bm r'),\label{complet}
\\
\frac{1}{4\pi^2}\int{d^3r\,\delta'(y-\bm n \cdot \bm r) \delta' ( y'-\bm n' \cdot \bm r)}=\delta(y-y')\delta(\bm n, \bm n')-\delta(y+y')\delta(\bm n, -\bm n').\label{ortonorma}
\end{gather}
We now review the properties of the tomographic transform of the scalar, the vector and the fermionic field.\\
Following \cite{Sommerfield:1982qf}, the tomographic transform of the scalar field is defined as:
\begin{equation}
\label{scalartransform}
\tilde \phi(y, \bm n) = \frac{1}{2\pi}\int{ d^3 r\,  \delta'(y-\bm n \cdot \bm r) \phi(\bm r)}=-\tilde \phi(-y,- \bm n),
\end{equation}
while the inverse tomographic transformation is:
\begin{equation}
\phi(\bm r)= \frac{1}{4\pi}\int{dy\, d^2 n\,  \delta'(y-\bm n \cdot \bm r)\tilde \phi(y, \bm n) }.
\end{equation}
For what concerns the vector field $A_{\mu}$, the tomographic transforms of its four spacetime components are organized as: 
\begin{gather}
\label{vectortransform1}
\tilde A^S (y, \bm n)=\frac{1}{2\pi}\int{d^3 r\, \delta'(y-\bm n \cdot \bm r) A_0 (\bm r)},
\\
\tilde A^L (y, \bm n)=\frac{1}{2\pi}\int{d^3 r\, \delta'(y-\bm n \cdot \bm r) \bm n \cdot \mathbf A (\bm r)},
\\
\label{vectortransform2}
\tilde A^{Ta} (y, \bm n)=\frac{1}{2\pi}\int{d^3 r\, \delta'(y-\bm n \cdot \bm r) \bm \varepsilon^a(\bm n) \cdot \mathbf A (\bm r)},
\end{gather}
where $\tilde A^L (y, \bm n)$ and $\tilde A^{Ta} (y, \bm n)$ are the longitudinal and the transverse transforms respectively, and we have introduced the polarization vectors $\bm \varepsilon^a (\bm n)$ orthogonal to $\bm n$, with $a=1,2$.
The spatial anti-transform is defined as:
\begin{equation}
\mathbf A(\bm r)=\frac{1}{4\pi}\int{dy\,d^2n \, \delta'(y-\bm n \cdot \bm r) \left[\bm n  \tilde A^L (y, \bm n) + \bm \varepsilon^a(\bm n)  \tilde A^{Ta} (y, \bm n)\right]}.\label{antivec}
\end{equation}
The tomographic transform of  the four component spinor field $\psi_\alpha(\bm r)$ is:
\begin{gather}
\tilde \psi^{b}(y, \bm n) = \frac{1}{2\pi}\int{ d^3 r\,  \delta'(y-\bm n \cdot \bm r) u_{\alpha}^{\dagger b}(\bm n) \psi_\alpha(\bm r)}, \label{spinnn}
\end{gather}
$u_{\alpha}^{\dagger b}$ is a spinor and where $\alpha = \{1,...,4\}$ is a spinor index. 
Introducing the usual $4\times4$ Dirac matrices 
$\bm \alpha$ and the spin matrices $\bm\Sigma= -\frac i2 \bm\alpha \times \bm\alpha$, $u_{\alpha}^{b}$ is, by definition, an eigenspinor of $\bm \alpha \cdot \bm n$ with eigenvalue $-1$. Moreover, since $[\bm\Sigma\cdot \bm n,\bm \alpha\cdot \bm n]=0$, $u_{\alpha}^{b}(\bm n)$ is also an eigenvector of $\mathbf \Sigma \cdot \bm n$, with:
\begin{gather}
(\mathbf \Sigma \cdot \bm n) u_{\alpha}^{b}(\bm n)=b u_{\alpha}^{b}(\bm n), \qquad b=\{1,-1 \}. \label{sigmau}
\end{gather}
The orthogonality condition $u_{\alpha}^{\dagger b}(\bm n) u_{\alpha}^{c}(\bm n)=\delta^{bc}$ holds, so we can write the projector as
\begin{equation}
\sum_{b} u_{\alpha}^{b}(\bm n)u_{\beta}^{\dagger b}(\bm n)=\frac12 (1-\bm\alpha \cdot \bm n)_{\alpha\beta}.\label{proiettore}
\end{equation}
The anti-transform of a spinor field is defined as
\begin{equation}
\psi_{\alpha}(\bm r)= \frac{1}{2\pi}\int{dy\, d^2 n\,  \delta'(y-\bm n \cdot \bm r)u_{\alpha}^{ b}(\bm n)\tilde \psi^{ b}(y, \bm n) } \label{antitomospin}
\end{equation}
Finally, by using the following identity:
\begin{equation}
\partial_i\delta'(y-\bm n \cdot \bm r)=-n_i\partial_y \delta'(y-\bm n \cdot \bm r),\label{fundamental}
\end{equation}
and the completeness relation \eqref{complet}, it is possible to prove for a massless Dirac field \eqref{spinnn}, {satisfying the equation $\gamma^\mu\partial_\mu\psi=0$}, in the tomographic representation requires to satisfies:
\begin{gather}
\left(\partial_0 - \partial_y\right)\tilde \psi^{b}(y, \bm n)=0.\label{dirac+}
\end{gather}
{Last equation shows that, for a fixed value of $\bf n$, $\psi^{b}(y, \bm n)$ is a``right moving'' field propagating along the positive direction of y.}
It is also possible to define the tomographic transform of a spinor field as follows:
\begin{equation}
\tilde \chi^{b}(y, \bm n) = \frac{1}{2\pi}\int{ d^3 r\,  \delta'(y-\bm n \cdot \bm r) v_{\alpha}^{\dagger b}(\bm n) \psi_\alpha(\bm r)}, \label{spinnn1}
\end{equation}
where $v_{\alpha}^{\dagger b}$ is an eigenspinor of $\bm \alpha \cdot \bm n$ with eigenvalue $+1$. For $v_{\alpha}^{ b}$ the following relations hold:
\begin{eqnarray}
(\mathbf \Sigma \cdot \bm n) v_{\alpha}^{b}(\bm n)=b v_{\alpha}^{b}(\bm n), \qquad b=\{1,-1 \}\\
\sum_{b} v_{\alpha}^{b}(\bm n)v_{\beta}^{\dagger b}(\bm n)=\frac12 (1+\bm\alpha \cdot \bm n)_{\alpha\beta}. \qquad \; \;
\end{eqnarray}
The anti-transform of the spinor field, in this case, is defined as:
\begin{equation}
\psi_{\alpha}(\bm r)= \frac{1}{2\pi}\int{dy\, d^2 n\,  \delta'(y-\bm n \cdot \bm r)v_{\alpha}^{ b}(\bm n)\tilde \chi^{ b}(y, \bm n) }, \label{antitomospin1}
\end{equation}
while the tomographic transformed Dirac equation for $\tilde{\chi}^{ b}(y, \bm n)$ is:
\begin{equation}
(\partial_0+\partial_y)\tilde{\chi}^{ b}(y, \bm n)=0 \label{dirac-}.
\end{equation}
{Anologously here we have a ``left moving'' field.}
As we shall see, the two previous construction of the tomographic transform of the fermionic field are completely equivalent.
\subsection{Tomographic duality}

The duality relation \eqref{duality} can be written in terms of the tomographic transformed fields $\tilde{\Lambda}$ and $\tilde{\xi}^L$ as follows:
\begin{gather}
\partial_0 \tilde \Lambda(y, \bm n) = \partial_y \tilde \xi^L (y, \bm n)\label{lita1},
\\
\partial_y \tilde\Lambda (y, \bm n)= \partial_0 \tilde \xi^L(y, \bm n)\label{lita2}.
\end{gather}
On the other hand, the longitudinal condition \eqref{roto} requires that the transverse components vanish, as well as their tomographic counterparts $\tilde \xi^{Ta}(y, \bm n)\equiv0$.
Consequently, we find that $\tilde \xi^{L}(y, \bm n)$ is the tomographic transform of the unique d.o.f. of the massless tensor $\phi_{\mu\nu}$, and from now on $\tilde \xi^L (y, \bm n) \equiv \tilde \xi (y, \bm n)$.

\subsection{Fermionization}

In the previous sections we defined the properties of the tomographic transform and we wrote the duality relation \eqref{duality} in terms of tomographic variables \eqref{lita1} and \eqref{lita2}. Now we implement the fermionization process following the steps described in \cite{Aratyn:1984jz,Aratyn:1983bg}.\\

Identifying $\tilde\Lambda (y, \bm n)$ and $\tilde\xi (y, \bm n)$ as the tomographic transforms of the bosonic d.o.f. on which the $3+1D$ theory described by the Lagrangian \eqref{lag5d4d} depends, we define a fermionic field as:
\begin{equation}
\tilde \Psi (y, \bm n)=\frac{1}{\sqrt{2\pi\alpha}}:e^{i\sqrt{\pi}\left[\tilde \xi (y, \bm n) + \tilde \Lambda (y, \bm n)\right]}:,\label{spinop}
\end{equation}
where $\alpha$ is a regularizing constant. The normal ordering prescription appearing in \eqref{spinop} is $\{\tilde\xi^{(-)},\tilde\Lambda^{(-)},\tilde\Lambda^{(+)},\tilde\xi^{(+)}\}$. The ordinary time evolution of the free fields $\Lambda$ and $\xi$
allows us to write down the Heisenberg operators of the positive and negative 
frequency parts of $\tilde\xi$ and $\tilde\Lambda$:
\begin{gather}
\tilde \xi^{(\pm)} (y, \bm n)=\int_{-\infty}^{\infty}{\frac{dp}{2\pi}\frac{1}{\sqrt{2|p|}}a^{(\pm)}(p) e^{\mp \left(ipy-i|p|t\right)}e^{-\frac{\alpha}{2}|p|}},
\\
\tilde \Lambda^{(\pm)} (y, \bm n)=\int_{-\infty}^{\infty}{\frac{dp}{2\pi}\frac{1}{\sqrt{2|p|}}b^{(\pm)}(p) e^{\mp \left(ipy-i|p|t\right)}e^{-\frac{\alpha}{2}|p|}},
\end{gather}
and $a^{(+)}(p)$ ($a^{(-)}(p)$) and $b^{(+)}(p)$ ($b^{(-)}(p)$) are the creation (annihilation) operators for $\tilde \xi (y, \bm n)$ and $\tilde \Lambda (y, \bm n)$ respectively. Because of the duality relations \eqref{lita1} and \eqref{lita2}, they are related by 
\begin{gather}
b^{(\pm)}(p)=-\mathrm{sign}(p)a^{(\pm)}(p)
\end{gather}
with $\mathrm{sign}(x)$ denotes the sign of $x$.
Let us define $\phi(y, \bm n)$ as:
\begin{equation}
\phi(y, \bm n)=\tilde \xi(y, \bm n) + \tilde \Lambda(y, \bm n),
\label{Phiright}
\end{equation}
and, accordingly,
\begin{equation}
 \phi^{(\pm)} (y, \bm n)=\int_{0}^{\infty}{\frac{dp}{\pi}\frac{1}{\sqrt{2|p|}}a^{(\pm)}(p) e^{\mp \left(ipy-i|p|t\right)}e^{-\frac{\alpha}{2}|p|}}.
\end{equation}
The presence of only positive $p$ momenta reminds what happens in the $1+1D$ 
bosonization, where only the right moving components are involved \cite{Shankar:1995rm}. The minus sign in \eqref{Phiright} would correspond to terms with only negative momenta 
components, in analogy with the left movers in $1+1D$.\\
The crucial observation is that the Lorentz scalar defined in \eqref{spinop} is a fermionic anticommuting variable. To see this, we compute the anticommutator 
\begin{multline}
\left\{\tilde \Psi (y, \bm n),\, \tilde \Psi^{\dagger} (y', \bm n)\right\}=\\
\frac{1}{2\pi\alpha}:e^{i\sqrt{\pi}\left[\phi^{(+)}(y, \bm n)+\phi^{(-)}(y, \bm n))\right]}: :e^{-i\sqrt{\pi}\left[\phi^{(+)}(y', \bm n)+\phi^{(-)}(y', \bm n))\right]}:+
\\
 \frac{1}{2\pi\alpha}:e^{-i\sqrt{\pi}\left[\phi^{(+)}(y', \bm n)+\phi^{(-)}(y', \bm n))\right]}: :e^{i\sqrt{\pi}\left[\phi^{(+)}(y, \bm n)+\phi^{(-)}(y, \bm n))\right]}:.
\label{anti}
\end{multline}
We observe  that\footnote{Using the identity \cite{Shankar:1995rm} :
\begin{equation}
e^A e^B=:e^{A+B}:e^{\left<AB+\frac{A^2+B^2}{2}\right>}.\label{normaorde}
\end{equation}}
\begin{multline}
:e^{i\sqrt{\pi}\left[\phi^{(+)}(y, \bm n)+\phi^{(-)}(y, \bm n)\right]}: :e^{-i\sqrt{\pi}\left[\phi^{(+)}(y', \bm n)+\phi^{(-)}(y', \bm n)\right]}:\ =\\
e^{\pi\left<\phi(y, \bm n)\phi(y', \bm n) - \frac{\phi^2(y, \bm n) + \phi^2(y', \bm n)}{2}\right>}=\frac{\alpha}{\alpha-i(y-y')}.
\end{multline}
The second term in the r.h.s. of \eqref{anti} can be treated in the same way,
with the outcome that:
\begin{eqnarray}
\left\{\tilde \Psi (y, \bm n),\, \tilde \Psi^{\dagger} (y', \bm n)\right\}
&=&
\left[\frac{\alpha}{\alpha-i(y-y')}+\frac{\alpha}{\alpha+i(y-y')}\right]\frac{1}{2\pi\alpha}\nonumber\\
&=&
\frac{\alpha}{\pi(\alpha^2+ (y-y')^2)}\underset{\alpha\to0}\to \delta(y-y').
\end{eqnarray} 

Therefore we verified that the scalar field defined in \eqref{spinop} satisfies the anticommutation relations for $\bm{n} =\bm{n'} $, but the commutation relation is still bosonic for $\bm{n} \neq\bm{n'} $. The complete Fermi statistics can be achieved introducing the Klein factors \cite{Luther:1978cy}:
\begin{equation}
\label{def1}
O_{\bm n}=e^{\frac{i\sqrt \pi}{2} \int^{(\theta, \varphi)}d^2 n'[\alpha(\bm n')+\beta(\bm n')]},
\end{equation}
where the operators
\begin{eqnarray}
\alpha(\bm n) &=& \int_{-\infty}^{\infty}{dy \,\partial_0 \tilde \xi(y,\bm n)} \\
\beta(\bm n) &=& \int_{-\infty}^{\infty}{dy \,\partial_0 \tilde \Lambda(y,\bm n)},
\label{def2}\end{eqnarray}
take the role of a generalized charges \cite{Aratyn:1984jz,Aratyn:1983bg}.
The angle parametrization $(\theta',\phi')$ define the versor direction $\bm n'$ and the integration domain $\int^{(\theta, \varphi)}$ is $0 < \varphi' \le 2\pi \cup 0 < \theta' < \theta $ if $\theta' \ne \theta$ and $0 < \varphi' < \varphi$ if $\theta' = \theta$.
Using the definitions \eqref{def1}-\eqref{def2}, the following rule can be verified 
\begin{equation}\label{invert}
O_{\bm n}\tilde \Psi(y, \bm m)=\begin{cases} -\tilde \Psi(y, \bm m)O_{\bm n} & \bm m < \bm n\\
\tilde \Psi(y, \bm m)O_{\bm n}& \bm m \ge \bm n\end{cases},
\end{equation}
where $\bm m < \bm n$ means $\theta_{\bm m} < \theta_{\bm n}$ or $\theta_{\bm m} = \theta_{\bm n}$ and $\varphi_{\bm m} < \varphi_{\bm n}$. \\
At this point it is straightforward to check that the operator
\begin{equation}
\tilde{\psi}(y, \bm n) \equiv \tilde{ \Psi} (y,\bm{n}) O_{\bm{n}}
\label{defferm}
\end{equation}
satisfies the anticommutation relations
\begin{equation}
\begin{split}
&\{\tilde \psi (x,\bm n),\, \tilde \psi^{\dagger} (y, \bm m)\}=\delta(\bm n, \bm m)\delta(x-y),
\\
&\{\tilde \psi (x,\bm n),\, \tilde \psi (y, \bm m)\}=0,
\end{split}
\end{equation}
which allow us to conclude that $\tilde\psi (y, \bm{n})$ is a well defined fermionic field in the tomographic representation.\\
Moreover, by using the definition \eqref{defferm} and the duality relations \eqref{lita1} and \eqref{lita2}, it is easy to see that $\tilde{\psi}$ must satisfy
\begin{equation}
\left(\partial_0 - \partial_y\right)\tilde \psi(y, \bm n)=0,\label{dirac--}
\end{equation}
which is the tomographic version of the massless Dirac equation \eqref{dirac+}.\\

An additional fermionic field can be introduced:
\begin{equation}
\tilde \chi (y, \bm n)=\frac{1}{\sqrt{2\pi\alpha}}:e^{i\sqrt{\pi}\left[\tilde \xi (y, \bm n) - \tilde \Lambda (y, \bm n)\right]}:P_{\bm n}, \label{chi}
\end{equation}
where
\begin{equation}
P_{\bm n}=e^{\frac{i\sqrt \pi}{2} \int^{(\theta, \varphi)}d^2 n'[\alpha(\bm n')-\beta(\bm n')]},
\end{equation}
which satisfies the Fermi statistics:
\begin{equation}
\begin{split}
&\{\tilde \chi (x,\bm n),\, \tilde \chi^{\dagger} (y, \bm m)\}=\delta(\bm n, \bm m)\delta(x-y),
\\
&\{\tilde \chi (x,\bm n),\,\tilde \chi  (y, \bm m)\}=0,
\end{split}
\end{equation}
together with the massless Dirac equation \eqref{dirac-}:
\begin{equation}
\left(\partial_0 + \partial_y\right)\tilde \chi(y, \bm n)=0.\label{dirac++}
\end{equation}
Consequently $ \tilde \chi(y, \bm n)$ obeys the tomographic construction \eqref{spinnn1}.\\
As we have anticipated, it is possible to reconstruct the spinor $\psi_{\alpha}(r)$ both from $ \tilde \psi(y, \bm n)$ and $ \tilde \chi(y, \bm n)$ in a completely equivalent way. In fact, choosing a fixed eigenvalue $b$ of $\mathbf \Sigma \cdot \bm n$ associated to $ \tilde \chi(y, \bm n)$ \footnote{We are not able do determine the value of $b$ a priori.}, we have from \eqref{spinnn1}:
\begin{multline}
\psi_{\alpha}(r)= \int dy d^2n \delta'(y- \bm n \cdot \bm r) v^b_{\alpha} (\bm n) \tilde \chi(y, \bm n)=\;(y \rightarrow -y, \; \bm n \rightarrow - \bm n)=\\
\int dy d^2n\delta'(-y+ \bm n \cdot \bm r) v^b_{\alpha} (-\bm n) \tilde \chi(-y, -\bm n). \label{cc}
\end{multline}
But, keeping in mind that $\delta'(y- \bm n \cdot \bm r)$ is an odd function under the transformation $(y \rightarrow -y, \; \bm n \rightarrow - \bm n)$ and the well known relation $v^b(- \bm n)=u^{-b}( \bm n)$ \footnote{Remember that $ \bm \alpha \cdot \bm n u^b(\bm n)=-u^b(\bm n)$ and $ \bm \alpha \cdot \bm n v^b(\bm n)=v^b(\bm n)$}, we obtain that \eqref{cc} is equal to
\begin{equation}
\label{dd}
- \int dy d^2n  \delta'(y- \bm n \cdot \bm r) u^{-b}_{\alpha} (\bm n) \tilde \chi(-y,- \bm n).
\end{equation}
Finally, since, by construction,  $\tilde \xi(-y,-\bm n)=\tilde \xi(y,\bm n)$ and $\tilde \Lambda(-y,-\bm n)=-\tilde \Lambda(y,\bm n)$, we obtain from the definition of $\tilde \chi(y, \bm n)$ \eqref{chi} that
\begin{equation}
\tilde \chi(-y,-\bm n)=\tilde \psi (y,\bm n),
\end{equation} 
and consequently \eqref{dd} is equal to
\begin{equation}
-\int dy d^2n \delta'(y- \bm n \cdot \bm r) u^{-b}_{\alpha}(\bm n) \tilde \psi (y, \bm n),
\end{equation}
which proves that the construction \eqref{spinnn} and \eqref{spinnn1} are completely equivalent.\\
Finally, we are dealing with only one independent tomographic fermionic field, from which we can only construct a Weyl spinor since, as it is well known, for massless fermion in 3+1D, $\mathbf \Sigma \cdot \bm n$ is equivalent to $\gamma^5$ and consequently our tomographic transformed spinor field is an eigenvalue of $\gamma^5$ by construction. Then, we can use it only to construct a Weyl spinor. 

\section{The integrated charge density}

In this section, as an example, we compute the integrated charge density expressed in terms of the tomographic transformed bosonic fields of our starting model.

The total charge, expressed in terms of the tomographic fermionic variables is:
\begin{equation}
\rho_F=\int d^3r\psi^{\dagger}_\alpha( \bm{r}) \psi_\alpha (\bm{r}) =
\int{dy \, d^2 n\,  \tilde\psi^{\dagger}(y, \bm n) \tilde\psi(y, \bm n)  },
\label{densita}
\end{equation}
where we have used the generalization of the ortonormality relation \eqref{ortonorma} for the fermionic field:
\begin{equation}
\frac{1}{4\pi^2}\int{d^3r\,\delta'(y-\bm n \cdot \bm r)\delta'(y'-\bm n' \cdot \bm r)u_{\alpha}^{\dagger b}(\bm n) u_{\alpha}^{c}(\bm n')}=\delta^{bc}\delta(y-y')\delta(\bm n, \bm n').
\end{equation}
We evaluate the r.h.s of \eqref{densita} with the point-splitting regularization technique \cite{Shankar:1995rm}:
\begin{multline}
\label{aa2}
\rho_{\psi}=\lim_{\varepsilon \to 0} \left[\tilde\psi^{\dagger}(y + \varepsilon, \bm n) \tilde\psi(y, \bm n)\right]=
\\
=\lim_{\varepsilon \to 0} \left[\frac{1}{2\pi}\left(\frac{1}{i\varepsilon}-\sqrt{\pi}\partial_y \left[\tilde \xi(y, \bm n) + \tilde \Lambda(y, \bm n)\right]\right)\right].
\end{multline}
Here we have used the usual point-splitting assumption, with the limit $\alpha\to0$ taken 
before than $\epsilon\to0$.\\
The total charge is obtained by subtracting to \eqref{aa2} the vacuum average charge \footnote{Actually the total charge is an integral of a total derivative and, as usual, is equal to zero if specific boundary conditions are not imposed on the fields of the theory.}:
\begin{multline}
\bar \rho_F= \rho_F-\left<\rho_F\right>=\lim_{\varepsilon \to 0} \int{dy \, d^2 n\,  \left[\frac{1}{2\pi}\left(\frac{1}{i\varepsilon}-\sqrt{\pi}\partial_y\left( \tilde \xi(y, \bm n)+\tilde \Lambda(y, \bm n) \right) - \frac{1}{i\varepsilon}\right)\right]}
\\=
-\int{dy \, d^2 n \, \frac{1}{2\sqrt{\pi}}\partial_y \left( \tilde \xi(y, \bm n)+\tilde \Lambda(y, \bm n) \right)}.
\end{multline}
{This result is exactly what we have expected, since $\tilde \psi (y, \bm n)$ is 
a tomographic ``right moving'' fermion because, in the representation of \eqref{spinnn}, it satisfies the 
tomographic Dirac equation with the minus sign \eqref{dirac+}.
Analogously the current associated with this spinor must also be ``right moving'' and indeed it only depends on the ``right moving'' combination of fields $\tilde \xi (y, \bm n)+ \tilde \Lambda (y, \bm n)$.}\footnote{{Equivalently in the representation \eqref{antitomospin1} only the negative $p$ 
components need to be considered and in such case we will obtain a minus sign between the 
two bosonic fields.}}

\section{Summary of results}

In this paper we explicitly constructed the fermionic d.o.f. for a $3+1D$ bosonic 
theory where a scalar field and a tensor field are related by a duality relation. The most natural 
interpretation of this duality relation comes from the dimensional reduction (on a planar boundary) of 
the $4+1D$ BF theory. This is done in complete analogy with the $3+1$ TI, where the $3+1D$ BF theory for the bulk, if restricted to the $2+1D$ boundary, naturally displays fermionic d.o.f. \cite{Cho:2010rk}. From a more field theoretical point of view, we stress that the duality relation is not imposed by hand, but emerges as the unique boundary condition for the fields of the topological $4+1D$ BF model with a planar boundary \cite{Amoretti:2013nv}. Following the tomographic representation of quantized fields presented in \cite{Sommerfield:1982qf}, we have given the tomographic representation of a fermionic field corresponding to the bosonic original d.o.f.. We have shown that this fermionic field satisfies the correct anticommutation relations and the massless Dirac equations. In addition, we have shown that it is only possible, with our tomographic transformed spinor field, to construct a Weyl spinor. Finally, as an explicit example, the fermionic integrated charge density has been considered, and we showed that its bosonized tomographic counterpart coincides, indeed, with what we expected for the right mover tomographic fermion we are dealing with.

\section{Acknowledgements}
We thank the support of INFN Scientific Initiative SFT: ``Statistical Field Theory, Low-Dimensional Systems, Integrable Models and Applications'' and FIRB - Futuro in Ricerca 2012" - Project HybridNanoDev RBFR1236VV. A.B. acknowledge the hospitality of the Institute for Nuclear Theory (INT-PUB-13-036) in Seattle where the work was partially done.


\appendix




\begin{thebibliography}{100}
\bibitem{Savit:1980sa} 
  R.~Savit,
  {\it ``Duality in field theory and statistical systems''},
  Rev.\ Mod.\ Phys.  {\bf 52}, 453 (1980)
\bibitem{Seiberg:1994rs} 
  N.~Seiberg and E.~Witten,
  {\it ``Electric - magnetic duality, monopole condensation, and confinement in N=2 supersymmetric Yang-Mills theory''},
  Nucl.\ Phys.\ B {\bf 426}, 19 (1994)
  [Erratum-ibid.\ B {\bf 430}, 485 (1994)]
  [hep-th/9407087].
\bibitem{Giveon:1994fu} 
  A.~Giveon, M.~Porrati and E.~Rabinovici,
  {\it ``Target space duality in string theory''}
  Phys.\ Rept.\  {\bf 244}, 77 (1994)
  [hep-th/9401139].
\bibitem{Witten:1998qj} 
  E.~Witten,
  {\it ``Anti-de Sitter space and holography''},
  Adv.\ Theor.\ Math.\ Phys.\  {\bf 2}, 253 (1998)
  [hep-th/9802150].
  \bibitem{Polchinski:2010hw} 
  J.~Polchinski,
  {\it ``Introduction to Gauge/Gravity Duality''},
  arXiv:1010.6134 [hep-th].
   \bibitem{Sachdev:2010ch} 
  S.~Sachdev,
  {\it ``Condensed Matter and AdS/CFT''}
  Lect.\ Notes Phys.\  {\bf 828}, 273 (2011)
  [arXiv:1002.2947 [hep-th]].
  \bibitem{Luttinger63} 
  J.~M.~Luttinger,
  {\it ``An exactly soluble model of a many fermion system''},
 J.\ Math.\ Phys.\   {\bf 4}, 1154 (1963).
\bibitem{MattisLieb65} 
  D.~C.~Mattis and E.~H.~Lieb,
  {\it ``Exact solution of a many fermion system and its associated boson Field  ''},
 J.\ Math.\ Phys.\   {\bf 6}, 304 (1965).  
\bibitem{Mattis75} 
  D.~C.~Mattis,
  ``{\it New wave - operator identity applied to the study of persistent currents in 1D''},
 J.\ Math.\ Phys.\   {\bf 15}, 609 (1974).
 \bibitem{Coleman75} 
  S.~R.~Coleman,
  {\it ``The quantum Sine-Gordon equation as the massive Thirring model''},
  Phys.\ Rev.\ D {\bf 11}, 2088 (1975).
\bibitem{Mandelstam75} 
  S.~Mandelstam,
  {\it ``Soliton operators for the quantized Sine-Gordon equation''}
  Phys.\ Rev.\ D {\bf 11}, 3026 (1975).
\bibitem{Haldane81} 
  F.~D.~M.~Haldane,
  {\it ``'Luttinger liquid theory' of one-dimensional quantum fluids. I. Properties of the Luttinger model and their extension to the general 1D interacting spinless Fermi gas '',}
  J.\ Phys.\ C.\  {\bf 14}, 2585 (1981).
  \bibitem{Shankar:1995rm} 
  R.~Shankar,
  {\it``Bosonization: How to make it work for you in condensed matter,''}
  Acta Phys.\ Polon.\ B {\bf 26}, 1835 (1995).
\bibitem{ShoellerDelft98} 
  J.~von~Delft and H.~Schoeller,
  {\it ``Bosonization for beginners and refermionization for experts'',}
  J.\ Phys.\ C.\  {\bf 14}, 2585 (1981).
  \bibitem{Luther:1978cy}
  A.~Luther,
  {\it``Bosonized fermions in three-dimensions''}
  Phys.\ Rept.\  {\bf 49}, 261 (1979).
\bibitem{Burgess:1994np} 
  C.~P.~Burgess and F.~Quevedo,
  {\it ``NonAbelian bosonization as duality''}
  Phys.\ Lett.\ B {\bf 329}, 457 (1994)
  [hep-th/9403173].
\bibitem{Chan:2012nb} 
  A.~Chan, T.~L.~Hughes, S.~Ryu and E.~Fradkin,
  {\it ``Effective field theories for topological insulators by functional bosonization''},
  Phys.\ Rev.\ B {\bf 87}, 085132 (2013)
  [arXiv:1210.4305 [cond-mat.str-el]].
\bibitem{Tsvelik92} 
  A.~M.~Tsvelik,
  {\it ``New fermionic description of quantum spin liquid state''},
  Phys.\ Rev.\ Lett. {\bf 69}, 2142 (1992)
\bibitem{Coleman93} 
  P.~Coleman, E.~Miranda and A.~M.~Tsvelik,
  {\it ``Possible realization of odd-frequency pairing in heavy fermion compounds''},
  Phys.\ Rev.\ Lett. {\bf 70}, 2960 (1993)
\bibitem{Sommerfield:1982qf} 
  C.~M.~Sommerfield,
  {\it ``Tomographic representation of quantized fields''}
  In *New Haven 1981, Proceedings, Symmetries In Particle Physics*, 127-140.
\bibitem{Aratyn:1984jz} 
  H.~Aratyn,
  {\it ``Fermions from bosons In (2+1)-dimensions''},
  Phys.\ Rev.\ D {\bf 28}, 2016 (1983).
\bibitem{Aratyn:1983bg} 
  H.~Aratyn,
  {\it ``A Bose representation for the massless Dirac field in four-dimensions''}
  Nucl.\ Phys.\ B {\bf 227}, 172 (1983).
\bibitem{HasanKane10} 
  M.~Z.~Hasan, C.~L.~Kane,
  {\it ``A Bose representation for the massless Dirac field In four-dimensions''},
  Rev.\ Mod.\ Phys.\  {\bf 82}, 3045 (2010).
\bibitem{QiZhang11} 
  X.-L.~Qi, S.-C.~Zhang,
  {\it ``Topological insulators and superconductors''},
  Rev.\ Mod.\ Phys.\  {\bf 83}, 1057 (2011).
\bibitem{Birmingham:1991ty} 
  D.~Birmingham, M.~Blau, M.~Rakowski and G.~Thompson,
  {\it ``Topological field theory''},
  Phys.\ Rept.\  {\bf 209}, 129 (1991).
\bibitem{Cho:2010rk} 
  G.~Y.~Cho and J.~E.~Moore,
  {\it ``Topological BF field theory description of topological insulators''},
  Annals Phys.\  {\bf 326}, 1515 (2011)
  [arXiv:1011.3485 [cond-mat.str-el]].
\bibitem{Blasi12} 
A.~Blasi, A.~Braggio, M.~Carrega, D.~Ferraro, N.~Maggiore and N.~Magnoli,
  {\it ``Three-dimensional dynamics of four-dimensional topological BF theory with boundary''},
  New J.\ Phys.\  {\bf 14}, 013060 (2012).  
\bibitem{Guadagnini90} 
E.~Guadagnini, N.~Maggiore and S.~P.~Sorella, 
  {\it ``Supersimmetry of the three-dimensional Einstein-Hilbert gravity in the Landau gauge''},
   Phys. Lett. B {\bf 247}, 543 (1990).  
\bibitem{Qi08} 
X.-L.~Qi, T.~L.~Hughes and S.-C.~Zhang, 
  {\it ``Topological field theory of time-reversal invariant insulators''},
   Phys. Rev. B {\bf 78}, 195424 (2008).  
\bibitem{Freedman12} 
D.~Z.~Freedman and A.~van~Proeyen, 
  {\it ``Supergravity''},
   Cambridge University Press, Cambridge (2012).     
\bibitem{Regge83} 
T.~Regge, 
  {\it ``The group manifold approach to unified gravity''},
   in Les Houches 1983, Proceedings, Relativity, Groups and Topology, Conf. Proc. {\bf C8306271}, 933 (2008).  
   \bibitem{Amoretti:2012hs} 
  A.~Amoretti, A.~Blasi, N.~Maggiore and N.~Magnoli,
  {\it ``Three-dimensional dynamics of four-dimensional topological BF theory with boundary''},
  New J.\ Phys.\  {\bf 14}, 113014 (2012).
\bibitem{Burkov11} 
A.~A.~Burkov and L.~Balents, 
  {\it ``'Weyl semimetal in a topological insulator multilayer'},
   Phys. Rev. Lett. {\bf 107},  127205 (2011).
\bibitem{Xu11} 
G.~Xu, H.~Weng, Z.~Wang, X.~Dai and Z.~Fang, 
  {\it ``'Chern Semimetal and the Quantized Anomalous Hall Effect in HgCr$ _2$Se$ _4$'},
  Phys. Rev. Lett. {\bf 107}, 186806 (2011). 
\bibitem{Amoretti:2013nv}
  A.~Amoretti, A.~Blasi, G.~Caruso, N.~Maggiore and N.~Magnoli,
  {\it``Duality and dimensional reduction of  BF theory"},
   Eur.\ Phys.\ J.\ C {\bf 73}, 2461 (2013)
  arXiv:1301.3688 [hep-th].
\bibitem{Diamantini:2011aa} 
  M.~C.~Diamantini and C.~A.~Trugenberger,
  {\it ``Charge spin separation in 3D''}
  arXiv:1112.3281 [cond-mat.str-el].
\bibitem{Kalb:1974yc}
  M.~Kalb and P.~Ramond,
  {\it``Classical direct interstring action,''}
  Phys.\ Rev.\ D {\bf 9},  2273 (1974).
\bibitem{Abhiraman95}
  R.~Abhiraman and C.~M.~Sommerfield,
  {\it``On abelian bosonization of free Fermi fields in three space dimensions''}
 arXiv:9501008 [hep-th]
\end{thebibliography}
\end{document}